\documentclass{article}

\usepackage{PRIMEarxiv}

\usepackage[utf8]{inputenc} 
\usepackage[T1]{fontenc}    
\usepackage{hyperref}       
\usepackage{url}            
\usepackage{booktabs}       
\usepackage{amsfonts}       
\usepackage{nicefrac}       
\usepackage{microtype}      
\usepackage{lipsum}
\usepackage{fancyhdr}       
\usepackage{graphicx}       
\graphicspath{{media/}}     
\usepackage{xcolor}
\usepackage[round]{natbib}
\usepackage{minted}
\usepackage{tikz}
\usepackage[inkscapelatex=false]{svg}
\usepackage{amsmath}
\usepackage{multirow}

\graphicspath{{fig/}}

\title{Advancing Subsurface Discovery and Geothermal Monitoring with an Agentic Artificial Intelligence Framework
}

\author{
  Randy Harsuko\textsuperscript{*1,2}, Zhengfa Bi\textsuperscript{1}, Guodong Chen\textsuperscript{1,2}, Nori Nakata\textsuperscript{1} \\
  \textsuperscript{1}Lawrence Berkeley National Laboratory, Berkeley, CA, USA \\
  \textsuperscript{2}University of California, Berkeley, CA, USA \\
  \textsuperscript{*}\texttt{\{mharsuko@lbl.gov\}}
}

\begin{document}
\maketitle

\begin{abstract}
Geothermal field development typically involves complex processes that require multi-disciplinary expertise in each process. Thus, decision-making often demands the integration of geological, geophysical, reservoir engineering, and operational data under tight time constraints. We present Geothermal Analytics and Intelligent Agent, or GAIA, an AI-based system for automation and assistance in geothermal field development. GAIA consists of three core components: GAIA Agent, GAIA Chat, and GAIA Digital Twin, or DT, which together constitute an agentic retrieval-augmented generation (RAG) workflow. Specifically, GAIA Agent, powered by a pre-trained large language model (LLM), designs and manages task pipelines by autonomously querying knowledge bases and orchestrating multi-step analyses. GAIA DT encapsulates classical and surrogate physics models, which, combined with built-in domain-specific subroutines and visualization tools, enable predictive modeling of geothermal systems. Lastly, GAIA Chat serves as a web-based interface for users, featuring a ChatGPT-like layout with additional functionalities such as interactive visualizations, parameter controls, and in-context document retrieval. To ensure GAIA’s specialized capability for handling complex geothermal-related tasks, we curate a benchmark test set comprising various geothermal-related scenarios and rigorously evaluate the system’s performance. Beyond task-level automation, GAIA is a unified framework that tightly couples physics-based modeling with agentic reasoning, which enables both domain-infused agentic evolutionary algorithm (meta-evolution) and end-to-end geothermal data processing within a single system. This integration highlights GAIA’s potential not only as an assistive tool but as a platform for accelerating scientific discovery and enabling more autonomous, data-driven geothermal field development.
\end{abstract}

\keywords{AI agents, geothermal, digital twin, scientific discovery, evolutionary algorithm}

\maketitle

\section{Introduction}
Geothermal energy is a renewable energy source that harnesses heat from the Earth's interior for electricity generation and direct use applications. It has the potential to provide a significant portion of the world's energy needs while reducing greenhouse gas emissions. However, geothermal field development is a complex process that involves multiple stages, including resource assessment, exploration, drilling, production, and reservoir management and optimization. Each stage requires specialized knowledge and expertise in various fields such as geology, geophysics, reservoir engineering, and environmental science \citep{gensheng2024current}. 
The complexity of geothermal field development presents challenges for decision-making, as it requires experts to integrate diverse data sources, interpret complex geological and geophysical information, and make informed choices under time constraints.
To address these challenges, there is a growing interest in leveraging artificial intelligence (AI) and machine learning (ML) techniques to automate and assist in geothermal field development (e.g., \cite{he2020injection, nakata2025ml, gutierrez2025ai, montesmaximizing}). Recently, AI agents have shown promise in various domains, including natural language processing (NLP), computer vision (CV), and robotics \citep{chan2024visibility}.

AI agents are LLM-powered systems that integrate dynamic reasoning, adaptive planning, multi-iteration external data retrieval and tool use, and comprehensive analytical report generation for research purposes \citep{huang2025deep}. They are meant to assist users in automating complex tasks, such as data analysis, literature review, and experimental design. Some examples of recent AI agents include OpenAI Deep Research \citep{OpenAI}, Alita \citep{qiu2025alita}, Pangu DeepDiver \citep{shi2025pangu}, etc. For a more comprehensive review of AI agents, please refer to \cite{huang2025deep}.

Beyond productivity gains, agentic systems have recently demonstrated potential to accelerate scientific discovery. Examples include AI-guided optimization of mathematical algorithms \citep{thind2025optimai}, automated hypothesis generation from large scientific corpora \citep{gottweis2026accelerating}, autonomous design of experiments \citep{schmidgall2025agent}, and the discovery of improved computational procedures through evolutionary search \citep{novikov2025alphaevolve}. By combining large-scale knowledge retrieval with iterative reasoning and tool execution, agentic systems can explore solution spaces that would be impractical for humans to investigate exhaustively. In geoscience and geothermal applications, such capabilities are particularly attractive because scientific discoveries often emerge from integrating heterogeneous observations, physical constraints, simulation outputs, and domain expertise across multiple disciplines. Agentic systems therefore provide a potential pathway toward more autonomous scientific workflows in which data interpretation, model formulation, algorithm development, and hypothesis testing are tightly coupled.

In the geoscience sphere, several LLM-based systems have been proposed. ChatClimate \citep{vaghefi2023chatclimate} is a web-based chatbot app built for querying climate-related information based on the latest IPCC report. It leverages the standard retrieval-augmented generation (RAG) pipeline to reduce hallucinations and provides more accurate and factual answers on climate. Pioneering LLM-based geological map analysis, \cite{huang2025peace} promote PEACE, an agentic system with GPT-4o as a base model that excels in various geological map-understanding tasks. 
They also curate a benchmarking dataset for evaluating AI agents in geologic map understanding. AskGDR \citep{weers2024empowering} is the first LLM-based app introduced to the geothermal community. The chatbot-styled web app was trained on metadata from collections of geothermal datasets in the Geothermal Data Repository (GDR) and was designed to assist researchers in obtaining geothermal datasets that suit their needs. For seismic exploration applications, \cite{kanfar2025intelligent} is the first to introduce an AI agent that specializes in building seismic processing workflows. Rather than manually writing Madagascar \citep{fomel2013madagascar} scripts, the agent can automatically generate and execute the scripts based on user prompts, significantly reducing the time and effort required for seismic data processing \citep{kanfar2025intelligent}. However, currently, there is still limited, if not any, research on agentic AI systems that handle a complete, end-to-end geothermal field development workflow.

The objective of this work is to develop an AI-based system, GAIA -- Geothermal Analytics and Intelligent Agent, for automation and assistance in geothermal field development. GAIA addresses the complexity and challenges of decision-making, which often requires integrating geological, geophysical, reservoir engineering, and operational data under time constraints. The system is designed to assist experts throughout the workflow, from data analysis and simulation to decision support and project automation. To the best of our knowledge, GAIA is the pioneering effort in building an agentic AI system for geothermal project assistance.

\section{Methods}
\begin{figure}
    \centering
    \includegraphics[width=0.7\textwidth]{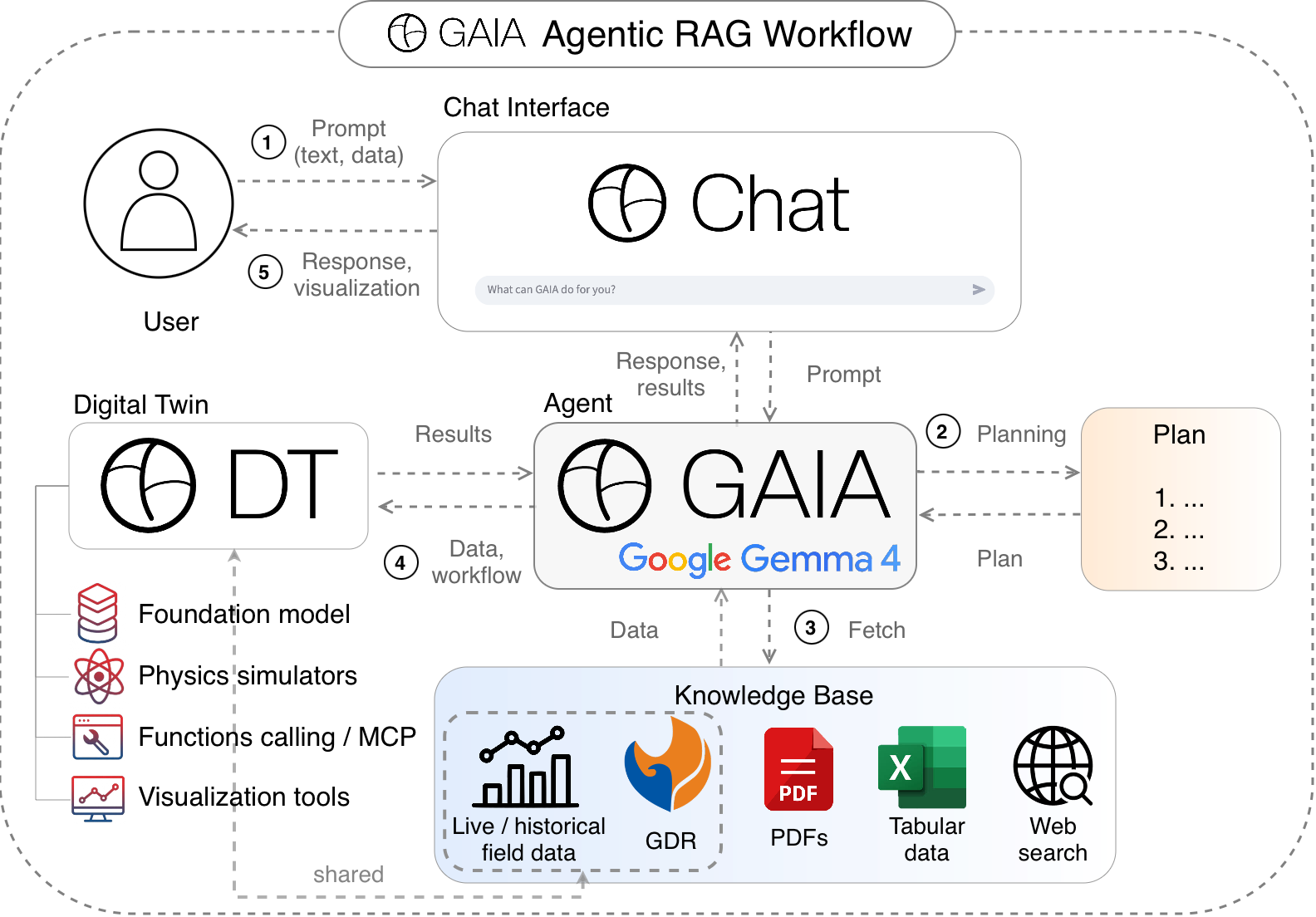}
    \caption{GAIA Agentic RAG Workflow. Users interact with the GAIA Chat interface by supplying prompts and supplementary data. The GAIA Main Agent will first think through the prompt and make a step-by-step plan to solve the given problem. If necessary, the system will fetch data from its knowledge base consisting of PDFs of papers/proceedings, tabular data, and/or external data sources (via web search and/or GDR). The collected data and the formulated workflow will be forwarded to GAIA DT for the actual processing through its set of tools. Finally, the Main Agent will formulate the final response based on the gathered information and multi-step analyses.
    }
    \label{fig:diagram}
\end{figure}

GAIA comprises three components: GAIA Agent, GAIA Chat, and GAIA Digital Twin or DT, forming an agentic RAG workflow. GAIA Agent, powered by a pre-trained LLM, autonomously queries knowledge bases, executes domain-specific subroutines, and orchestrates multi-step analyses. GAIA DT integrates classical and surrogate physics models with interactive visualization tools, while GAIA Chat provides a web-based interface with features such as interactive visualizations, parameter controls, and context-aware document retrieval. Figure \ref{fig:diagram} illustrates the GAIA Agentic RAG workflow. Each of these components will be discussed in detail next.

\subsection{GAIA Agent}
\label{sec:gaia_agent}
The core component of agentic AI systems is an AI agent, which is based on a pre-trained LLM. In the GAIA framework, we define a main agent that is responsible for planning, orchestrating, and executing tasks. Therefore, the minimum requirement of the pre-trained LLM is that it needs to understand instructions (e.g., via instruction tuning \citep{peng2023instruction}) and generate human-like text. In this work, we use the open-weight model of Gemma 4 from Google Deepmind \citep{team2025gemma}, which is a 26B parameters Mixture-of-Experts (MoE, \cite{shazeer2017outrageously}) LLM model. We specifically use the Gemma4-26B-A4B-IT (instruction-tuned) variant of the model, which is designed to follow user instructions and generate coherent and contextually relevant responses. We also tested the previous version of Gemma (i.e., Gemma 3) and compared their performance in answering capability, which are discussed in Appendix~\ref{sec:rag_benchmarks}.

We acknowledge that other LLMs can be used as the main agent, such as OpenAI's GPT-5, Meta's Llama 4, and Anthropic's Claude 4. We choose Gemma 4 for its open-weight availability and its strong performance in instruction-following tasks, comparable to one of the earlier flagship LLMs, Gemini \citep{team2023gemini, team2024gemini}. Moreover, smaller LLMs can easily fit into consumer-grade GPUs, making it more accessible for users to run GAIA on their local machines and edge devices in case there is no internet connection (such as in remote geothermal fields).

The main agent operates in a combination of the Chain-of-Thought (CoT, \cite{wei2022chain}) and Reaction+Act (ReAct, \cite{yao2023react}) paradigms. Specifically, the main agent will first think through the prompt and make a step-by-step plan to solve the given problem. In each step, the main agent will decide whether to deduce from the knowledge base or execute tools via the GAIA DT. Then, for every step, the main agent will analyze whether the step has been completed and the objective of the step has been achieved. These processes are done iteratively until the main objective, which is to answer the user's prompt, is achieved. The main agent will then formulate the final response based on the gathered information and the multi-step analyses.

If necessary, GAIA Agent will fetch data from its knowledge base consisting of PDFs of papers/proceedings, tabular data, and/or external data sources (via web search and/or GDR). We leverage the RAG paradigm \citep{lewis2020retrieval} to enhance the agent's ability and efficiency in retrieving relevant information from the knowledge base. The key to the RAG pipeline is an embedding model, which is responsible for converting multi-modal data (text, images, etc.) into vector representations. A strong embedding model should be capable of capturing the semantic meaning of the data and generating high-quality embeddings that can be used for similarity search. In this work, we use the open-weight model of Jina v4 \citep{gunther2025jina}, the latest open-weight embedding model with multi-modal (text and images) support. The embedding model is used to convert the multi-modal data into vector representations, which are then stored in a vector database for efficient retrieval.

\paragraph{GAIA RAG system} Originally, RAG systems were designed to handle text data only \citep{lewis2020retrieval}. First, texts are tokenized and converted into vector representations using an embedding model. The vector representations are then stored in a vector database, which allows for efficient similarity search and retrieval. When a user submits a query, the query is also converted into a vector representation using the same embedding model. The vector database is then queried to find the most similar texts to the query based on their vector representations using a similarity metric (e.g., cosine similarity). The retrieved texts are then used as context for the LLM to generate a response. However, in geothermal field development, data comes in various formats, including text documents (research papers, reports, etc.), images (figures, maps, etc.), and tabular data (CSV files, spreadsheets, etc.). Therefore, we extend the traditional RAG system to handle multi-modal data by leveraging a multi-modal embedding model. For an efficient retrieval process, we store all the vector representations in a single vector database, along with their metadata to distinguish between different data types.

For the internal knowledge base, we have collected over 5,000 papers and proceedings related to geothermal field development. Prior to embedding, we preprocess the documents by splitting them into smaller chunks of text (equivalent to 2,048 tokens) using Docling \citep{livathinos2025docling} to ensure that the embedding model can effectively capture the semantic meaning of each chunk. The chunked texts are then converted into vector representations using the embedding model and stored in a vector database (we use LanceDB \footnote{ https://lancedb.github.io/lancedb/}). For images, we use the same embedding model to convert the images into vector representations, which are also stored in the vector database. For tabular data, we treat them as images and convert them into vector representations using the embedding model. Whenever the main agent needs to retrieve information from the knowledge base, it will query the vector database using the embedding model to find the most relevant chunks of text and images. We filter the retrieved contents with a reranker model to ensure that only the most relevant information is used for response generation. The retrieved chunks will then be used as context for the main agent to generate a response. In the future, we plan to extend the knowledge base to include more data sources, such as web search results and GDR data, to further enhance the agent's ability to retrieve relevant information. 

We tested the GAIA RAG system on a set of geothermal-related queries and found that it can effectively retrieve relevant information from the knowledge base and generate accurate and informative responses. Compared to its baseline model, GAIA Agent with RAG capabilities provides up to 3.1x improvement in the measured metrics. For more details on the evaluation of the GAIA RAG system, please refer to the Appendix \ref{sec:rag_benchmarks}.

\begin{figure}
    \centering
    \includegraphics[width=1\textwidth]{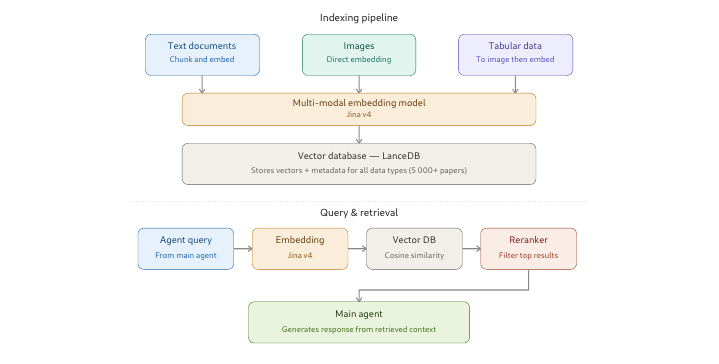}
    \caption{GAIA RAG system architecture. The system consists of a multi-modal embedding model that converts text, images, and tabular data into vector representations. These vector representations are stored in a vector database, which allows for efficient retrieval based on similarity search. When the agent submits a query, it is also converted into a vector representation, and the vector database is queried to find the most similar data. The retrieved data is then used as context for the agent to generate a response.}
\end{figure}

\subsection{GAIA DT}
GAIA DT is regarded as a digital twin in a broad sense, encompassing not only physics-based simulations but also data processing pipelines and AI-assisted workflows that connect modeling to real-world field operations and multi-modal diverse datasets. It integrates classical and surrogate physics models with domain-specific subroutines and visualization tools to enable predictive modeling of geothermal systems. Designed as a modular and extensible platform, GAIA DT allows users to incorporate new models, data streams, and subroutines as needed. The current version of GAIA DT includes the following components:
\begin{itemize}
    \item Seismic waveform analysis: GAIA DT includes a set of tools for seismic waveform analysis, which can be used for tasks such as phase picking, event location estimation, magnitude estimation, and seismicity forecasting. These tools are mainly based on the ObsPy \citep{beyreuther2010obspy} library, which is a widely used open-source library for seismic data analysis. The module can handle various seismic data formats, such as SAC, SEED, and MiniSEED. We plan to incorporate more advanced processing tools in the next versions of GAIA DT, such as ML-based tools \citep{nakata2025ml} and real-time processor \citep{nakata2024microseismicity}.
    \item Visualization tools: Since there are many visualization types in geothermal field development, we let an AI agent that specializes in producing visualizations handle these tasks. Specifically, we use a variant of the Gemma 3 model that is fine-tuned on coding tasks, called CodeGemma. We then prompt the CodeGemma model to generate Python code for visualizations based on the main agent's \ref{sec:gaia_agent} requirements. The generated code will then be executed internally to produce the desired visualizations. The visualizations are based on Plotly \footnote{https://plotly.com}, a popular Python library for creating interactive plots and dashboards.
\end{itemize}

Although we realize that many other components need to be added to GAIA DT, we first choose to focus on seismic waveform analysis and visualization tools in the current version of GAIA. This is because seismic monitoring is a crucial aspect of geothermal field development, as it provides valuable information about the subsurface structure and dynamics of geothermal reservoirs. Moreover, seismic data is often complex and requires specialized knowledge and expertise to analyze and interpret. In the future, we plan to extend GAIA DT to include more components, such as geophysical inversion, reservoir simulation models, risk assessment, and economic analysis. Nevertheless, the key feature of GAIA DT is the autonomous selection of the provided tools and their parameters within, which are both tedious and critical tasks, especially in a fast-paced environment of geothermal monitoring. Through a correct prompting, users could also manually interfere and modify the parameters of the selected tools at their convenience.

\subsection{GAIA Chat}
User interface (UI) is an important component in AI agent systems, as it serves as the primary point of interaction between users and the AI agent. A well-designed UI can enhance user experience, facilitate effective communication, and enable users to leverage the capabilities of the AI agent more efficiently. Therefore, we developed GAIA Chat, a web-based interface that allows users to interact with GAIA Agent and GAIA DT seamlessly.

For ease of testing, prototyping, and deployment, we utilize Streamlit \footnote{https://streamlit.io} to build a web-based app for GAIA. Streamlit is a Python library that simplifies the creation of interactive web applications, particularly for data science and machine learning projects. It allows users to transform Python scripts into shareable web apps with minimal effort and without requiring front-end web development experience (HTML, CSS, JavaScript) and contains various templates and built-in widgets. For more information on Streamlit, please refer to \cite{khorasani2022web} or their website (\href{https://streamlit.io}{streamlit.io}). Note that the GAIA Chat interface can be ported to production-grade frameworks like JavaScript in the future for more flexible customization, but the concept of the features will be the same, as explained below.

\begin{figure}
    \centering
    \includegraphics[width=0.7\linewidth]{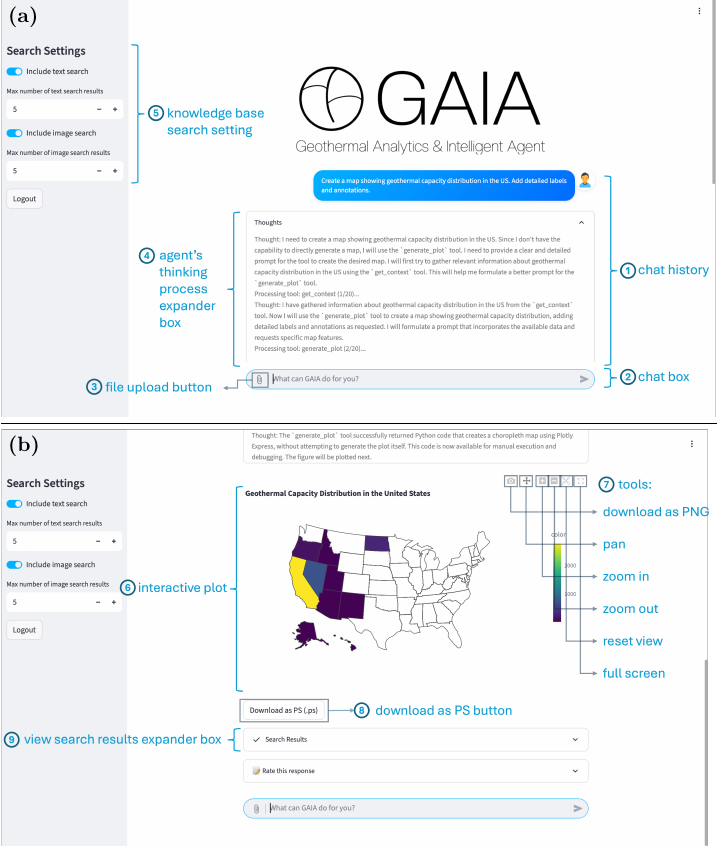}
    \caption{GAIA Chat interface. The main window features a chatbox to type and send messages, a conversation window to display message history, a file upload button, and a search setting panel to control the number of returned search items from the knowledge base. The interactive figure output allows users to pan, zoom, and hover over the generated figures, and export them to PNG or PostScript (PS).}
    \label{fig:gaia_chat}
\end{figure}

We adopt a layout similar to the ChatGPT web/desktop app (Figure \ref{fig:gaia_chat}). 
The main window of the GAIA web interface features components of a standard chatbot, like a chatbox to type and send messages, and a conversation window to display message history between a user and GAIA. Additionally, to support thorough geothermal analytics, we added the following features:
\begin{itemize}
    \item File upload button: supports any file extensions to account for a diverse type of geothermal data.
    However, currently, we focus on support for structured tabular data (like CSV) and common waveform file types (MSEED, SAC, SEG-Y, etc.). Note that users could also upload files via drag-and-drop to the chatbox.
    \item Search setting panel: to control the number of returned search items from the knowledge base. We allow users to separately choose the maximum number of documents and images for better flexibility.
    \item Interactive figure output: most figures displayed by GAIA will be interactive. Thanks to Plotly, a widget-focused Python plotting library, users can pan, zoom, and hover over the generated figures. Users can also export the figures to PNG or PostScript (PS) according to their needs.
\end{itemize}

\section{Applications}

In this section, we present the core contributions of GAIA to geothermal field development workflows. We emphasize how GAIA enables the integration of physics-based modeling and data-driven automation across various tasks. These contributions can be broadly categorized into (1) physics-aware inversion and (2) automated geothermal data processing and monitoring.

Traditional software packages and analysis pipelines can already perform many individual tasks discussed here, such as seismic processing, event location, or reservoir simulation. The distinguishing feature of GAIA is not necessarily a higher accuracy of the underlying algorithms themselves, but the autonomous orchestration, integration, and adaptation of these algorithms within a unified scientific workflow. This distinction is important because geothermal field development is increasingly constrained by the complexity of multidisciplinary decision-making rather than by the availability of individual analysis tools. By combining retrieval-augmented reasoning, domain knowledge, and physics-based modeling, GAIA can automatically formulate workflows, select tools, retrieve relevant scientific context, and iteratively refine solutions. Such capabilities reduce manual intervention, shorten analysis cycles, and enable more systematic exploration of alternative hypotheses.

\subsection{Physics-aware inversion}

Inverse modeling of subsurface fracture systems from thermal observations is a fundamentally ill-posed problem. Multiple fracture configurations can produce nearly identical temperature responses, particularly when observations are limited to scalar production temperature or sparse temperature-depth profiles. In practical geothermal workflows, this non-uniqueness is typically addressed through the incorporation of geological priors and physics-based constraints \citep{ceci2025geophysical}. 

In this section, we design a simplified, physics-inspired problem to demonstrate GAIA's capability in assisting the discovery of inverse algorithms under physical constraints. The objective is not to solve a realistic reservoir model, but to isolate the role of domain knowledge in guiding inverse problem formulation and solution.

\paragraph{Forward model}
We define a reduced-order forward model that relates fracture network properties to production temperature:
\begin{equation}
T_{\mathrm{prod}} = T_{\mathrm{res}} - \Delta T \left(1 - \exp(-C)\right),
\end{equation}
where $T_{\mathrm{res}}$ is the reservoir temperature, $\Delta T$ is the maximum thermal drawdown, and $C$ is a connectivity index defined as
\begin{equation}
C \propto n \left(\frac{b}{b_{\mathrm{ref}}}\right)^3 \left(\frac{d_{\mathrm{ref}}}{d}\right).
\end{equation}
Here, $n$ denotes the number of hydraulically active fracture zones (dimensionless), $d$ is the mean fracture spacing (m), and $b$ is the hydraulic aperture (m). The cubic dependence on aperture is motivated by the classical cubic law for fracture flow \citep{witherspoon1980validity}, while the dependence on fracture density and spacing reflects connectivity effects commonly modeled in discrete fracture network (DFN) studies \citep{cacas1990modeling, mi2017enhanced}. The exponential form captures saturation behavior observed in lumped geothermal reservoir models \citep{sarak2005lumped}.

\paragraph{Inverse problem formulation}
Given an observed production temperature $T_{\mathrm{obs}}$, the goal is to infer fracture parameters $\mathbf{m} = (n, d, b)$. This inverse problem is non-unique because different combinations of parameters can yield the same connectivity index $C$, and therefore the same $T_{\mathrm{prod}}$. Though we acknowledge that this problem under-represents real world scenario, it possesses similar non-uniqueness nature of most actual cases. 

To address this, we define an objective function combining data misfit and a plausibility term:
\begin{equation}
\label{eq:objective}
\mathcal{L} = \underbrace{|T_{\mathrm{pred}} - T_{\mathrm{obs}}|}_{\text{data misfit}} + \underbrace{\mathcal{P}(n, d, b)}_{\text{plausibility penalty}} + \underbrace{|\hat{\mathbf{m}}-\mathbf{m}|}_{\text{parameter misfit}} ,
\end{equation}
where $\mathcal{P}$ encodes prior knowledge on fracture properties and $\hat{\mathbf{m}}$ is the estimated parameter vector from the inverse algorithm.

\paragraph{Plausibility modeling}
The plausibility term is formulated as a normalized deviation from physically reasonable parameter distributions:
\begin{equation}
\mathcal{P}(n, d, b) = 
\left(\frac{n - \mu_n}{\sigma_n}\right)^2 +
\left(\frac{\log d - \mu_d}{\sigma_d}\right)^2 +
\left(\frac{\log b - \mu_b}{\sigma_b}\right)^2,
\end{equation}
where $(\mu, \sigma)$ denote representative mean and spread values derived from geothermal studies. For example, fracture spacing in enhanced geothermal systems can span $\sim 0.3$--$300$ m, while hydraulic apertures typically range from micrometers to sub-millimeter scales \citep{witherspoon1980validity,national1996rock}. This formulation corresponds to a maximum a posteriori (MAP) estimate, where the plausibility term acts as a prior over fracture parameters.

\paragraph{Algorithm-level optimization with AlphaEvolve}
Rather than directly optimizing a single fracture parameter vector $\mathbf{m}$, we use AlphaEvolve \citep{novikov2025alphaevolve} to optimize an inverse algorithm. AlphaEvolve is one example of a broader class of evolutionary AI systems that combine large language models with evolutionary search. Similar concepts have appeared in automated machine learning, program synthesis, neural architecture search, and scientific optimization (e.g., \cite{thind2025optimai, brookes2025evolving, gottweis2026accelerating}). We select AlphaEvolve because it provides an explicit framework for evolving executable algorithms through iterative mutation, evaluation, and selection.

Let $a$ denote an executable inverse algorithm that maps observations to fracture parameters:
\begin{equation}
a: T_{\mathrm{obs}} \mapsto \hat{\mathbf{m}}.
\end{equation}
Given a benchmark set of synthetic inverse problems
\begin{equation}
\label{eq:metaevolution_benchmark_dataset}
\mathcal{D} = \{T_{\mathrm{obs}}^{(i)}\}_{i=1}^{N},
\end{equation}
the quality of an inverse algorithm is measured by its expected loss across the benchmark:
\begin{equation}
J(a;\mathcal{D},E) =
\frac{1}{N}
\sum_{i=1}^{N}
E\left(a(T_{\mathrm{obs}}^{(i)}),T_{\mathrm{obs}}^{(i)}\right),
\end{equation}
where $E$ is the evaluator. In this example, the evaluator is defined as
\begin{equation}
E(\hat{\mathbf{m}},T_{\mathrm{obs}}) = w_T S_T + w_P S_P + w_{\theta} S_{\theta},
\end{equation}
where
\begin{equation}
S_T = \frac{1}{1 + \bar{e}_T},
S_P = \frac{1}{1 + \bar{e}_P},
S_{\theta} = \frac{1}{1 + \bar{e}_{\theta}},
\end{equation}
are the temperature score, plausibility score, and parameter score, respectively, and $\bar{e}_T$, $\bar{e}_P$, and $\bar{e}_{\theta}$ are the first, second, and third terms in Equation \ref{eq:objective}, respectively. The weights $w_T$, $w_P$, and $w_{\theta}$ are set to be equal in this example, but they can be adjusted to reflect different priorities in the optimization.

Let $a_0$ be the initial naive inverse algorithm, such as random search. AlphaEvolve applies a sequence of program transformations $\tau_1,\tau_2,\ldots,\tau_K$ to generate improved algorithms:
\begin{equation}
a_k = \tau_k(a_{k-1};\phi,E),
\end{equation}
where $\phi$ is the evolutionary parameters and $E$ is the evaluator. The AlphaEvolve optimization can therefore be written as
\begin{equation}
a^{\star}(\phi,E)
=
\arg\min_{a \in \mathcal{A}(a_0,\phi)}
J(a;\mathcal{D},E),
\end{equation}
where $\mathcal{A}(a_0,\phi)$ denotes the set of algorithms reachable from the initial program $a_0$ under the search behavior induced by AlphaEvolve parameters $\phi$.

\paragraph{GAIA-guided meta-evolution}
In the vanilla AlphaEvolve, the evolutionary parameters $\phi$ and evaluator $E$ are manually specified and remain fixed. In contrast, GAIA performs an outer-loop optimization over the context that guides algorithm evolution. Specifically, GAIA proposes a set of evolutionary initialization parameters $\phi$ and evaluator configuration $E_{\phi}$. For this simple example, we restrict the learnable parameters $\phi$ to be the system prompt and the evaluation algorithm. The system prompt is responsible for guiding the search behavior of AlphaEvolve, while the evaluation algorithm defines the criteria that AlphaEvolve optimizes.
The GAIA+AlphaEvolve procedure is then formulated as the bi-level optimization problem:
\begin{equation}
\phi^{\star}
=
\arg\min_{\phi}
J\left(
a^{\star}(\phi,E_{\phi});
\mathcal{D}_{\mathrm{val}},
E_{\phi}
\right),
\end{equation}
subject to
\begin{equation}
a^{\star}(\phi,E_{\phi})
=
\arg\min_{a \in \mathcal{A}(a_0,\phi)}
J(a;\mathcal{D}_{\mathrm{train}},E_{\phi}).
\end{equation}
The inner loop corresponds to AlphaEvolve improving the inverse algorithm, while the outer loop corresponds to GAIA improving the optimization context by modifying the prompt and evaluator priors.

Conceptually, this outer-loop optimization transforms the role of the agent from a workflow executor into a scientific assistant capable of shaping the discovery process itself. Instead of only searching for optimal model parameters, GAIA searches for improved optimization strategies, evaluation criteria, and domain-informed algorithm. Such meta-level optimization has the potential to accelerate the development of new scientific methodologies, particularly in complex geothermal problems where prior knowledge and physical intuition play a critical role.

\begin{table}
\caption{Metrics of all experiment results. The combined score is computed as an equally weighted summation of all other scores. The bold values represent the best values across all experiments. Note that AlphaEvolve yields the best temperature score due to overfitting to the temperature misfit.}
\centering
\small
\begin{tabular}{|c|c|c|c|c|c|}
\hline
\textbf{Setup} & \textbf{Temperature Score} & \textbf{Plausibility Score} & \textbf{Parameter Score} & \textbf{Combined Score} \\ 
\hline
Initial algorithm & 0.7790 & 0.1339 & 0.5444 & 0.6142 \\
\hline
AlphaEvolve & \textbf{0.9007 $\pm$ 0.0040} & 0.1398 $\pm$ 0.0113 & 0.5559 $\pm$ 0.0137 &  0.6491 $\pm$ 0.0050 \\
\textbf{GAIA + AlphaEvolve} & 0.8950 & \textbf{0.1520} & \textbf{0.5662} & \textbf{0.6532} \\
\hline
\end{tabular}
\label{tab:inversion_results}
\end{table}

\begin{figure}
    \centering
    \includegraphics[width=1\textwidth]{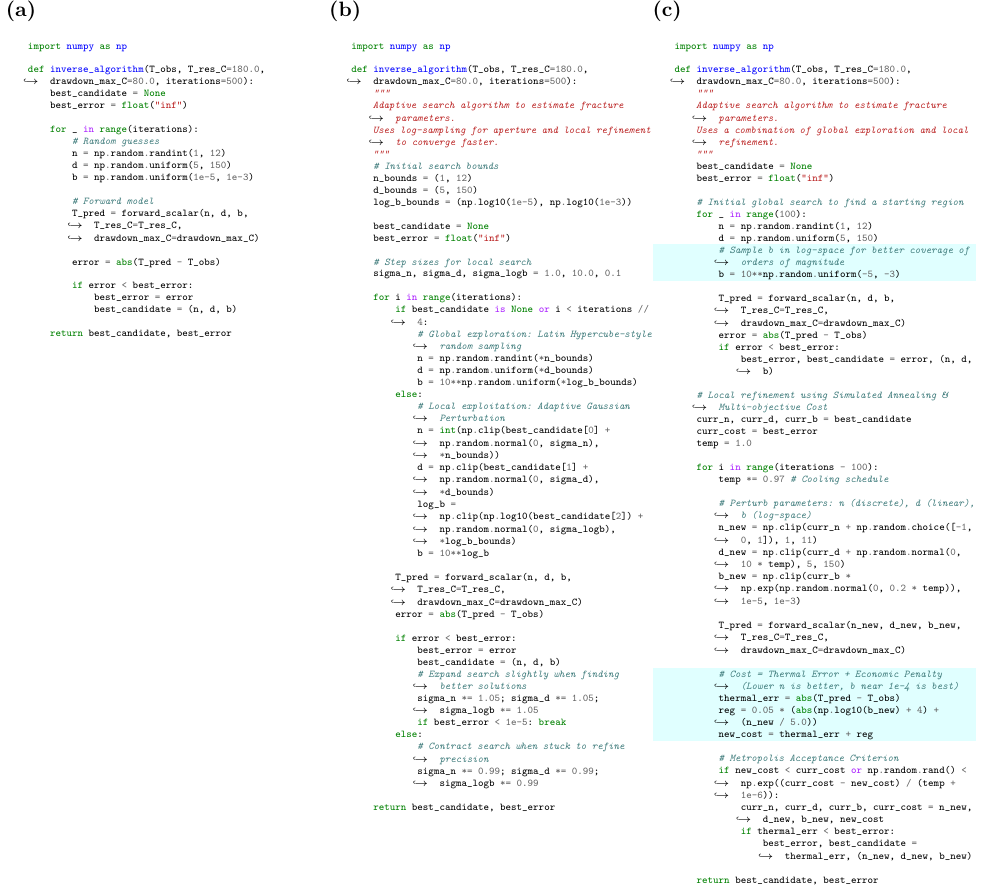}
    \caption{Comparison of the inversion algorithm from: (a) the initial state; (b) AlphaEvolve, and; (c) GAIA + AlphaEvolve. The highlighted lines in (c) represent the significant code snippets that induce geothermal domain knowledge to the problem.}
    \label{fig:inverse_algorithm}
\end{figure}

\begin{figure}
    \centering
    \includegraphics[width=0.7\textwidth]{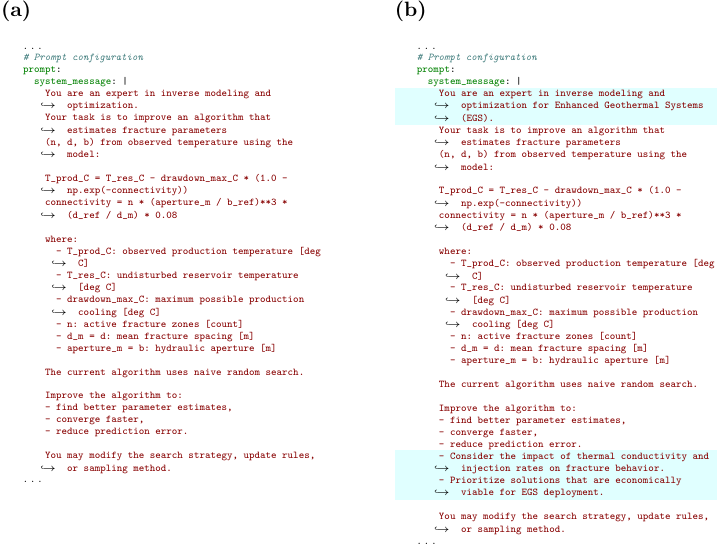}
    \caption{The initial system prompt (a) and the optimized system prompt (b) by the GAIA + AlphaEvolve framework. The highlighted lines in (b) represent the additional domain-knowledge information imposed by GAIA.}
    \label{fig:prompt_comparison}
\end{figure}

\paragraph{Experiments \& results}
We employ \textit{OpenEvolve} \citep{openevolve}, an open-source implementation of AlphaEvolve \citep{novikov2025alphaevolve}, to evolve an inverse algorithm that maps $T_{\mathrm{obs}}$ to fracture parameters. The initial algorithm consists of a naive random search over parameter space (Figure \ref{fig:inverse_algorithm}a), which is iteratively improved through evolutionary operations.

Two configurations are compared:
\begin{itemize}
    \item \textbf{AlphaEvolve baseline}: We fix the system prompt and the evaluator across 10 independent runs of AlphaEvolve optimization. The prompt contains the problem definition and generic instructions to improve the algorithm, but does not include specific geothermal domain knowledge or constraints (Figure \ref{fig:prompt_comparison}a).
    \item \textbf{GAIA + AlphaEvolve}: We let GAIA optimize the system prompt and the evaluator configuration for a maximum of 10 AlphaEvolve iterations. We explicitly instruct GAIA to use its RAG capabilities (Section \ref{sec:gaia_agent}) to retrieve relevant geothermal domain knowledge and incorporate it into the prompt and evaluation criteria. Additionally, we tell GAIA to use the last best algorithm for each iteration to ensure improvement continuity.
\end{itemize}
Additionally, in both setups, we add random Gaussian noise to the synthetic dataset (Equation \ref{eq:metaevolution_benchmark_dataset}) to increase the problem difficulty, encourage the discovery of more robust algorithms, and avoid collapse to analytical solutions.

Table \ref{tab:inversion_results} summarizes the results of the experiments. Both AlphaEvolve and GAIA + AlphaEvolve significantly improve the initial algorithm, with GAIA + AlphaEvolve achieving the best overall performance across all metrics except for the temperature score, where AlphaEvolve slightly outperforms due to overfitting to the temperature misfit. This is attributed to the unconstrained solutions proposed by AlphaEvolve (see Figure \ref{fig:inverse_algorithm}b for an example taken from the best AlphaEvolve run), which can achieve low temperature error but yield implausible fracture parameters due to over-focusing on local refinement of the temperature score. In contrast, GAIA + AlphaEvolve discovers a more balanced algorithm that incorporates domain knowledge to constrain the search towards more plausible and generalizable solutions (Figure \ref{fig:inverse_algorithm}c). The optimized system prompt by GAIA (Figure \ref{fig:prompt_comparison}b) includes additional geothermal-specific information and constraints, which likely contributed to the improved plausibility and parameter scores. The proposed solution by GAIA + AlphaEvolve even considers economic viability (Figure \ref{fig:inverse_algorithm}b and Figure \ref{fig:prompt_comparison}b), which is a critical aspect in practical geothermal applications but is often overlooked in purely data-driven optimization. However, we noticed that GAIA + AlphaEvolve did not optimize the evaluator, likely because the evaluator design space is more complex and less intuitive than the system prompt, which is more directly interpretable and modifiable by the LLM. Hence, we omit the comparison of the evaluator in this experiment. Future work will explore more structured approaches to evaluator optimization, such as modular evaluators to dissect the problem or hierarchical evaluation frameworks.

This simplistic experiment illustrates a potential future role of agentic AI in scientific research. The primary outcome is not the recovered fracture parameters themselves, but the emergence of a modified inversion methodology that embeds geothermal-specific reasoning. Such modifications can be viewed as potential scientific discoveries because they encode new problem-solving strategies derived from the interaction between domain knowledge, physical constraints, and evolutionary optimization. While the present example is intentionally simplified, the same framework could be extended to more realistic inverse problems involving reservoir simulation, induced seismicity forecasting, or coupled thermo-hydraulic-mechanical processes.

\subsection{Automated geothermal data processing and monitoring}

Another growing line of research in geothermal field development is the use of robust, automated monitoring systems to track seismicity and reservoir behavior in near real-time. Seismic and reservoir monitoring are crucial for understanding subsurface processes, assessing reservoir performance, and ensuring operational safety. However, the high volume of seismic and reservoir data generated in geothermal fields can overwhelm traditional manual processing workflows. GAIA can address this challenge by automating key monitoring tasks through the GAIA Agent and GAIA DT capabilities. We demonstrate the application in four tasks: seismic phase picking, event location estimation, magnitude estimation, and seismicity forecasting.

\subsubsection{Seismic waveform phase picking}

Seismic event detection and phase picking are key components in geothermal seismicity monitoring. Conventionally, amplitude-based methods dominate the seismic event detection process, which is often followed by manual phase picking. However, these methods are often time-consuming and labor-intensive, especially in geothermal fields with high seismicity rates. GAIA can automate this process by integrating seismic waveform data analysis and phase picking algorithms into its workflow.

There are two phase picking methods currently implemented in GAIA:
\begin{itemize}
    \item The STA/LTA (Short-Term Average/Long-Term Average) method, which is a widely used trigger-based method for seismic event detection \citep{allen1978automatic}. It calculates the ratio of short-term to long-term average amplitudes of seismic waveforms to identify potential seismic events, and;
    \item The machine learning-based phase picking method, which utilizes a pre-trained model to classify seismic waveforms into different phases (P-wave and S-wave, see \cite{nakata2025ml}). This method can be more accurate and efficient than traditional STA/LTA methods, especially in complex seismic environments. Moreover, unlike STA/LTA, this method does not rely on input parameters, such as the short-term and long-term window lengths, which can be difficult to determine in practice.
\end{itemize}

\begin{figure}
    \centering
    \includegraphics[width=0.6\linewidth]{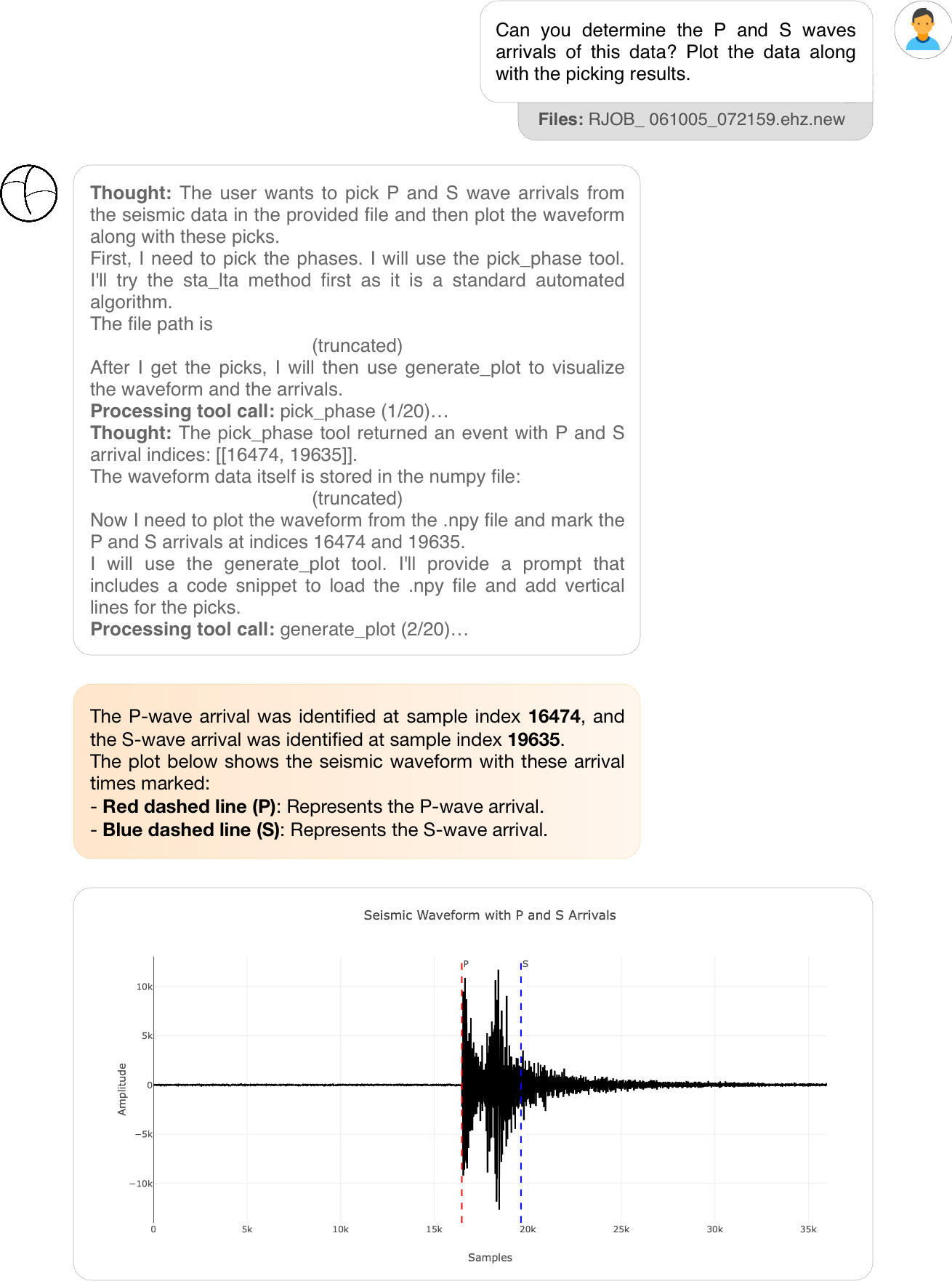}
    \caption{Example of GAIA's seismic waveform phase picking workflow. The user uploads a seismic waveform file and requests phase picking. GAIA Agent will then analyze the waveform data, apply the selected phase picking method, and return the results in an interactive plot.
    }
    \label{fig:phase_picking}
\end{figure}

The GAIA Agent can autonomously decide which method to use based on the characteristics of the seismic data and the user's preferences provided by GAIA Chat. The default picking function is the STA/LTA method, as it is more widely used and easier to interpret. The user can also choose to override the default method by specifying the desired phase picking method in the prompt. Figure \ref{fig:phase_picking} shows an example of GAIA's phase picking workflow, where the user uploads a seismic waveform file and requests phase picking. GAIA Agent will then analyze the waveform data, apply the selected phase picking method, and return the results in an interactive plot.

\subsubsection{Event location estimation}
\label{sec:event_location}

Event location estimation is a crucial step in seismic monitoring, as it determines the spatial coordinates of seismic events based on the detected phases. GAIA can automate this process by integrating various location estimation algorithms, such as the double difference method (HypoDD, \cite{waldhauser2001hypodd}) and the non-linear location method (NonLinLoc, \cite{lomax2000probabilistic,lomax2014earthquake}). These methods utilize the detected phases and their arrival times to calculate the event's location in three-dimensional space.

For simplicity, we currently implement the Geiger method \citep{geiger1912probability} as the default location estimation method in GAIA. The Geiger method is a linear location method that uses the arrival times of P- and S-waves to estimate the event's location. It is widely used in seismic monitoring due to its simplicity and effectiveness. However, in the future, we will add more options (like HypoDD and NonLinLoc), and users can choose other location estimation methods by specifying the desired method in the prompt.

\begin{figure}
    \centering
    \includegraphics[width=0.6\linewidth]{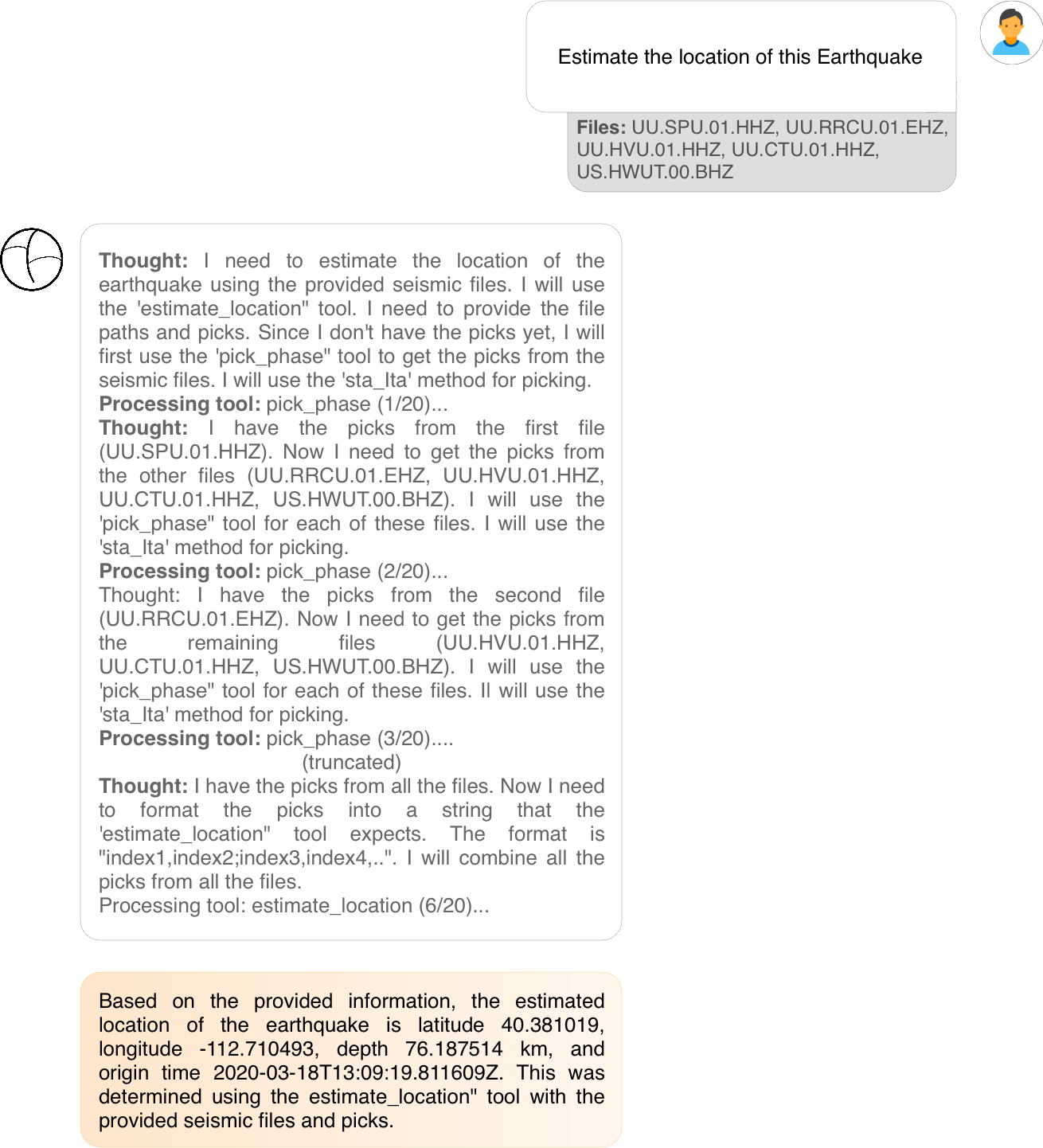}
    \caption{Example of GAIA's event location estimation workflow. The user uploads a seismic waveform file and requests event magnitude estimation. GAIA Agent will then analyze the waveform data and apply the selected location estimation methods.}
    \label{fig:event_location}
\end{figure}

Figure \ref{fig:event_location} shows an example of GAIA's event location estimation workflow, where the user uploads a seismic waveform file and requests event location estimation. GAIA Agent will then analyze the waveform data, apply the selected location estimation method, and return the location (latitude, longitude, and depth) and the origin time of the event.

\subsubsection{Event magnitude estimation}

Event magnitude estimation is another important aspect of seismic monitoring, as it quantifies the energy released by a seismic event. GAIA can automate this process by integrating various magnitude estimation algorithms, such as the Richter scale \citep{richter1935instrumental} and the Moment Magnitude (Mw) scale \citep{hanks1979moment}. These methods utilize the detected phases and their amplitudes to calculate the event's magnitude.

\begin{figure}
    \centering
    \includegraphics[width=0.6\linewidth]{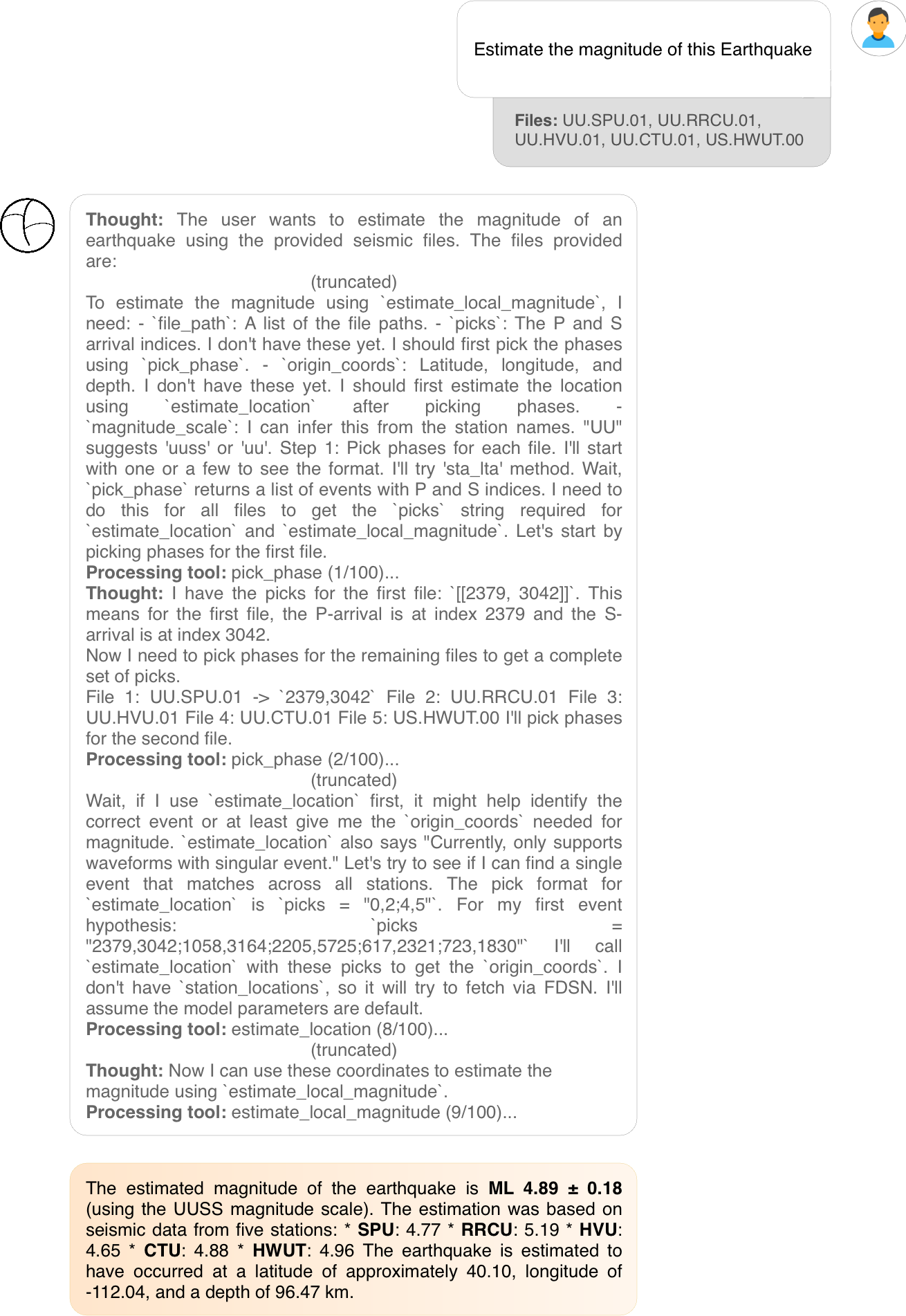}
    \caption{Example of GAIA's event magnitude estimation workflow. The user uploads a seismic waveform file and requests event magnitude estimation. GAIA Agent will then analyze the waveform data and apply the selected magnitude estimation methods.}
    \label{fig:event_magnitude}
\end{figure}

In the current version of GAIA, we implement the Richter scale, also known as the local magnitude (ML), as the default magnitude estimation method. The Richter scale is a logarithmic scale that measures the amplitude of seismic waves and is widely used in seismic monitoring. However, it has limitations in accurately estimating the magnitude of large events and events with complex waveforms. Moreover, the local magnitude parameters, such as the distance from the station to the event and the station correction factor, are often difficult to determine in practice. Therefore, we will add more options (like Moment Magnitude), and users can choose other magnitude estimation methods by specifying the desired method in the prompt.

Figure \ref{fig:event_magnitude} shows an example of GAIA's event magnitude estimation workflow, where the user uploads a seismic waveform file and requests magnitude estimation. Since the magnitude estimation requires the event's location, GAIA Agent will first estimate the location (Section \ref{sec:event_location}). Then, GAIA will apply the selected magnitude estimation method. The final response will be the event's magnitude.

\subsubsection{Seismicity forecasting}

Seismicity forecasting is a challenging task that aims to predict the occurrence of future seismic events based on historical data. Recent advances in machine learning have also demonstrated the potential of data-driven seismicity forecasting in geothermal systems. For example, \cite{biquakecastnet} proposed deep learning approaches for forecasting induced seismicity by learning from temporal patterns of earthquake catalogs. Such models can complement traditional statistical frameworks, like the Epidemic Type Aftershock Sequence (ETAS) model \citep{ogata1988statistical} and the Declustering method \citep{reasenberg1985second}, by capturing nonlinear relationships that may not be explicitly represented in physics-inspired formulations. Within GAIA, these forecasting approaches can be integrated as interchangeable modules, allowing the agent to select, compare, and evaluate different forecasting paradigms according to the available data and operational objectives.

\begin{figure}
    \centering
    \includegraphics[width=0.6\linewidth]{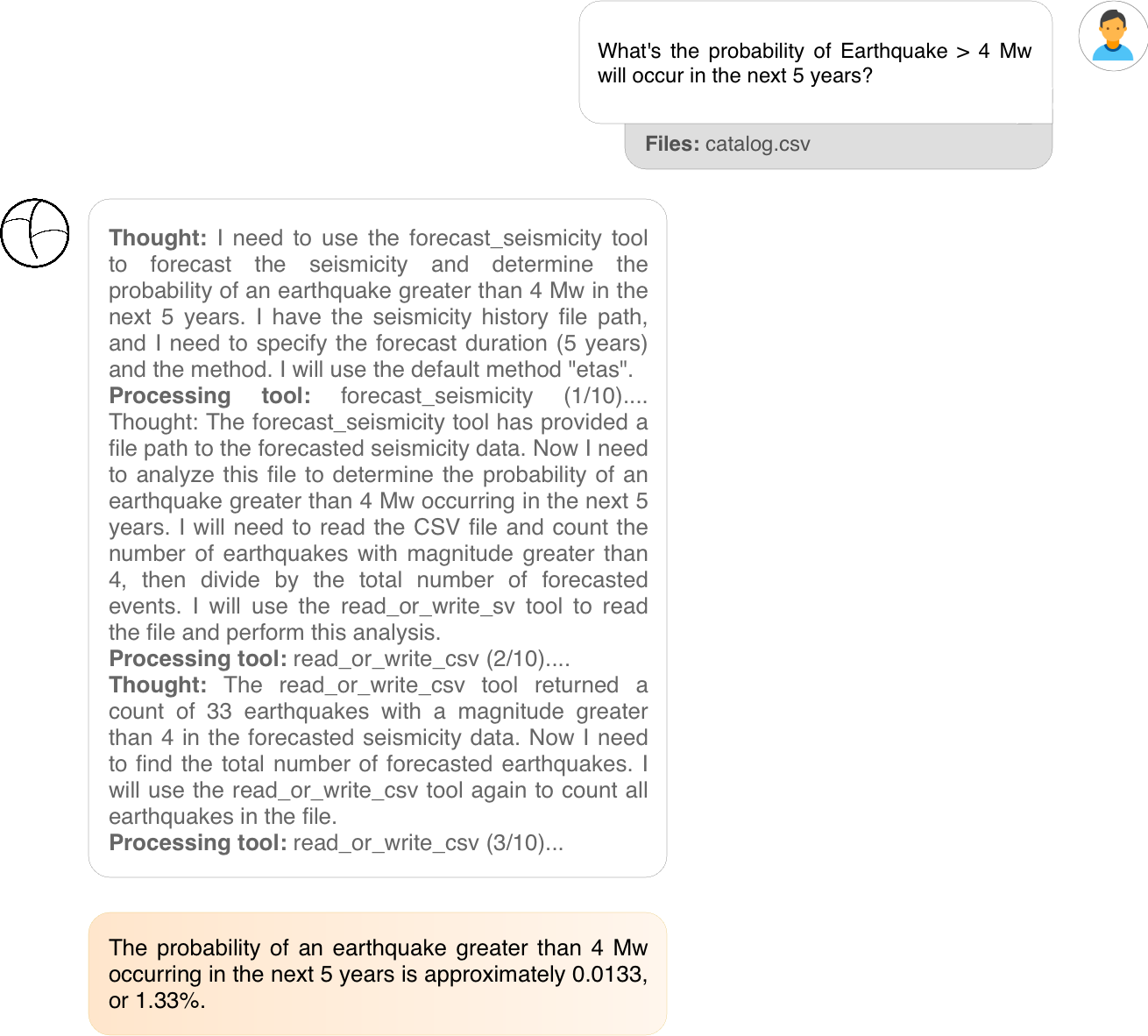}
    \caption{Example of GAIA's seismicity forecasting workflow. The user uploads a known seismicity catalog to the chatbox. GAIA Agent will then analyze the catalog and apply the selected forecasting model.}
    \label{fig:seismicity_forecasting}
\end{figure}

By default, GAIA uses the ETAS model as the seismicity forecasting method. The ETAS model is a statistical model that describes the temporal and spatial distribution of seismic events and is widely used in seismic monitoring. The upcoming version of GAIA will feature more advanced forecasting models, primarily based on deep learning techniques (e.g., \cite{bi2025advancing}). Figure \ref{fig:seismicity_forecasting} shows an example of GAIA's seismicity forecasting workflow, where the user uploads a known seismicity catalog and requests seismicity forecasting. GAIA Agent will then analyze the catalog, apply the selected forecasting algorithm, and return the forecasting results.

\section{Conclusions and Future Work}
GAIA represents a pioneering application of agentic RAG workflows in geothermal field development. Unlike conventional decision-support systems, it combines autonomous task orchestration, physics-based digital twin modeling, and interactive user engagement within a single platform. This integrated approach offers a scalable pathway toward intelligent, data-driven geothermal project management and operational automation.

Due to the modular design of GAIA and the fast-evolving AI agent ecosystem, we envision several avenues for future enhancements:
\begin{itemize}
    \item \textbf{Domain-focused LLMs}: While we currently use a general-purpose LLM (Gemma 4) as the main agent, we plan to explore domain-focused LLMs that are fine-tuned on geothermal-specific data. This could enhance GAIA's understanding of geothermal concepts and improve its ability to generate accurate and relevant responses. An example of such is GeoGPT \citep{huang2025geogpt}, which is a series of open-weight LLMs fine-tuned on geoscience-related data.
    \item \textbf{MCP tools integration}: Model Context Protocol (MCP) is a standardized open-source framework that enables interaction of AI systems with external applications. By integrating MCP tools into GAIA, we can expand its capabilities to perform various geothermal data analysis tasks without needing to build the tools from scratch. Some examples of MCP servers include: QGIS MCP \footnote{https://mcpmarket.com/server/qgis-model-context-protocol}, which is a QGIS \citep{QGIS_software} MCP server, for analyzing spatial data of geothermal fields; OSDU MCP \footnote{https://github.com/rasheedonnet/osdu-mcp}, which is Open Subsurface Data Universe (OSDU) MCP server, for accessing and managing open subsurface data stored in OSDU data platforms; and many more. Currently, there are no geothermal-specific MCP tools, but we are confident that the geothermal community will develop such tools in the near future. Therefore, we also plan to develop an MCP server for GAIA DT to allow seamless integration with other MCP-compatible applications.
    \item \textbf{Surrogate modeling and foundation models}: Physics-based simulations are often computationally expensive and time-consuming, which can limit their applicability in real-time geothermal monitoring and decision-making. To address this issue, we plan to incorporate surrogate modeling techniques into GAIA DT. Surrogate models are simplified representations of complex physics-based models that can approximate their behavior with significantly reduced computational costs. By integrating surrogate models into GAIA DT, we can enable faster simulations and analyses, allowing for real-time decision-making and optimization in geothermal field development. Moreover, we plan to explore the use of foundation models, which are large pre-trained models that can be fine-tuned for specific tasks. Foundation models have shown great promise in various domains, including natural language processing (NLP), computer vision (CV), and scientific computing. By leveraging foundation models in GAIA DT, we can enhance its ability to model complex geothermal systems and improve its predictive capabilities.
    \item \textbf{Self-evolving multi-agent systems}: The current version of GAIA uses a single main agent to orchestrate the workflow and interact with users. However, as geothermal field development involves various complex tasks and processes, we plan to explore the use of self-evolving multi-agent systems in GAIA. In a multi-agent system, multiple agents can collaborate and communicate with each other to achieve common goals \citep{li2024survey}. Each agent can specialize in specific tasks, such as data analysis, modeling, or visualization, and can work together to provide comprehensive solutions for geothermal field development. By implementing a self-evolving multi-agent system, we can enable GAIA to adapt and evolve its capabilities over time, allowing it to handle more complex tasks and improve its performance in geothermal project workflows.
\end{itemize}

Looking beyond the individual enhancements discussed above, our long-term vision for GAIA is the development of an autonomous scientific platform for geothermal energy systems. In such a framework, AI agents would continuously integrate new observations, retrieve relevant scientific knowledge, formulate and test hypotheses, perform physics-based simulations, and propose improved analytical methodologies. Rather than functioning solely as a decision-support tool, GAIA could become an active participant in the scientific process, accelerating the cycle of observation, interpretation, modeling, and discovery.

For geothermal field development, this capability could enable real-time adaptation of monitoring strategies, rapid identification of emerging reservoir behaviors, automated evaluation of physics, and continuous refinement of predictive models. More broadly, we envision GAIA as a prototype for future agentic scientific systems that tightly couple domain knowledge, physical modeling, and autonomous reasoning to accelerate discoveries in Earth and energy sciences.

\section*{Acknowledgments}
This work is partly supported by the U.S. Department of Energy's Hydrocarbons and Geothermal Energy Office under Contract No. DE-AC02-05CH11231 with Lawrence Berkeley National Laboratory (LBNL). We would also like to thank the Geothermal team at LBNL for their valuable feedback and contributions to the development of GAIA.

\bibliographystyle{abbrvnat}
\bibliography{references}  

\newpage

\appendix
\section{Appendix A: RAG Benchmarks}
\label{sec:rag_benchmarks}

To evaluate GAIA's performance, we curated a benchmark test set comprising various geothermal-related questions. The benchmark includes questions that require GAIA to retrieve information from its knowledge base, perform multi-step analyses, and generate comprehensive responses. The questions are generated using the Retrieval Augmented Generation Assessment (RAGAS) framework \citep{es2024ragas}, which provides a systematic approach to evaluate the performance of RAG systems by generating sets of questions with their corresponding ground truth contexts and answers from a given knowledge base. The benchmark covers a wide range of geothermal topics, including geothermal energy production, reservoir engineering, geophysical methods, and environmental impacts. We also added human-generated questions from experts in the geothermal field to ensure the benchmark's relevance and applicability. Some examples of the benchmark questions can be found in Table \ref{tab:rag_benchmark}.

We asked GAIA, equipped with either Gemma 3 or Gemma 4, to answer the benchmark questions and compared its performance with a baseline models (Gemma 3 and Gemma 4 without RAG capabilities). For a quantitative evaluation, we used the RAGAS framework to evaluate GAIA's performance in terms of response relevancy, response faithfulness, and response accuracy. We also used G-Eval \citep{liu2023g} to evaluate the correctness of the answers generated by GAIA and the baseline model. The core idea of both methods is to use a strong LLM as an evaluator to compare the generated responses with the ground truth answers and contexts, and assign scores based on various criteria, which is often referred to as LLM-as-a-judge method. The evaluation is done in a zero-shot manner, where the evaluator is given the generated response, the ground truth answer, and the ground truth context, and is asked to assign scores based on the criteria defined by RAGAS and G-Eval. We measure the following metrics:
\begin{itemize}
    \item Response accuracy: the correctness of the generated response compared to the ground truth answer.
    \item Response relevancy: the relevance of the generated response to the user's query
    \item Response faithfulness: the consistency of the generated response to the retrieved context
    \item Answer correctness: similar with response accuracy, but with a more comprehensive reasoning process, which is evaluated by G-Eval.
\end{itemize}

We chose Gemini 3.1 model \citep{team2023gemini, team2024gemini} as the evaluator for both RAGAS and G-Eval evaluations, as it is a strong Google's flagship LLM that can effectively evaluate the quality of the generated responses. The results of the evaluation can be found in Table \ref{tab:rag_performance}. GAIA outperforms the baseline models in all metrics, demonstrating its superior ability to retrieve relevant information from its knowledge base and generate accurate responses. In particular, we obtained up to 1.5x improvement in response accuracy, up to 1.3x improvement in response relevancy, up to 3.1x in response faithfulness, and up to 1.4x improvement in answer correctness compared to the baseline models.

\begin{table}[!t]
\caption{Examples of benchmark questions for GAIA's RAG performance evaluation.}
\centering
\small
\begin{tabular}{|c|p{0.25\textwidth}|p{0.32\textwidth}|p{0.32\textwidth}|}
\hline
\textbf{No} & \textbf{Question} & \textbf{Ground Truth Context} & \textbf{Ground Truth Answer} \\ 
\hline

1 & Could you elaborate on the methodology behind TFI and its applications within the Oil and Gas sector? 
& TFI is a proven method for direct 4D mapping of transmissive fracture/fault networks with over 50 successful projects. It uses Seismic Emission Tomography (SET) to capture weak but coherent seismic emissions from fracture zones. 
& TFI maps transmissive fracture/fault networks using SET by capturing coherent seismic emissions from hydraulically active zones. \\ 
\hline

2 & What is the typical procedure for measuring the magnitude of minimum principal stress in deep formations? 
& The minimum principal stress is typically measured using hydraulic fracturing-based tests such as DFIT, minifrac, or microfrac. 
& A pressurized fluid is injected into an isolated wellbore section to initiate fractures. Pressure fall-off is analyzed to estimate stress magnitude. \\ 
\hline

3 & What factors can cause deviations in geothermometer accuracy? 
& Conventional geothermometers may deviate due to fluid mixing, pressure variations, and non-equilibrium conditions. 
& Deviations arise from fluid mixing, pressure effects, and violations of equilibrium assumptions. \\ 
\hline

\end{tabular}
\label{tab:rag_benchmark}

\vspace{10pt}

\caption{Comparison of GAIA's RAG performance with the baseline model (Gemma 3 and Gemma 4) on benchmark questions. The table shows RAGAS metrics (response accuracy, response relevancy) and G-Eval metric (answer correctness).
}
\centering
\begin{tabular}{|l|p{2cm}|p{2cm}|p{2cm}|p{2cm}|}
\hline
\multirow{2}{*}{\textbf{Model}} 
& \multicolumn{3}{c|}{\textbf{RAGAS}} 
& \textbf{G-Eval} \\ \cline{2-4}
& \textbf{Response Accuracy} 
& \textbf{Response Relevancy} 
& \textbf{Response Faithfulness} 
& \textbf{Answer Correctness} \\ \hline

GAIA w/ Gemma 3 & 0.41 & \textbf{0.90} & \textbf{0.95} & 0.38 \\ \hline
GAIA w/ Gemma 4 & \textbf{0.48} & \textbf{0.90} & \textbf{0.95} & \textbf{0.46} \\ \hline
Baseline (Gemma 3) & 0.30 & 0.67 & 0.31 & 0.37 \\ \hline
Baseline (Gemma 4) & 0.32 & 0.76 & 0.47 &  0.32 \\ \hline
\hline
Improvement (GAIA w/ Gemma 3) & 1.4x & 1.3x & 3.1x & 1.0x \\ \hline
Improvement (GAIA w/ Gemma 4) & 1.5x & 1.2x & 2.0x & 1.4x \\ \hline

\end{tabular}
\label{tab:rag_performance}
\end{table}

\end{document}